\documentclass[twocolumn]{revtex4-1}
\usepackage{graphicx}
\usepackage{color}
\newcommand{\had}{\hat{a}^+}
\newcommand{\hadp}{\hat{a}^{\prime +}}
\newcommand{\hbdp}{\hat{b}^{\prime +}}

\newcommand{\hbd}{\hat{b}^+}

\newcommand{\vac}{|{\rm vac}\rangle}
\newcommand{\ket}[1]{\left| #1 \right\rangle}

\newcommand{\eea}{\end{eqnarray}}
\newcommand{\bea}{\begin{eqnarray}}
\newcommand{\ee}{\end{equation}}
\newcommand{\be}{\begin{equation}}

\begin{document}
\title{Thermalizing two identical particles}

\author{S.J. van Enk}

\affiliation{Department of Physics and
Oregon Center for Optical, Molecular \& Quantum Sciences\\
University of Oregon, Eugene, OR 97403}

\begin{abstract}
How do indistinguishable identical bosons manage to obey Bose-Einstein statistics---and hence be correlated---even when they do not interact with each other? Part of the answer is that the bosons have to interact {\em indirectly} with each other by interacting with the same environment. A joint measurement interaction provides a good example.
Thermalization occurs whenever there are two competing processes, one diagonal in the energy basis (namely, reversible Hamiltonian evolution), the other irreversible  and diagonal in a complementary basis (for example, a measurement in a spatially localized basis). Correlations arise only from initial states in which the bosons start in different (orthogonal) states.

\end{abstract}

\maketitle
\subsection{Question}
Two identical indistinguishable bosons follow  Bose-Einstein (B-E) statistics.
The difference between B-E statistics and 
classical Maxwell-Boltzmann  statistics lies in how we count the number of states in which one boson is in a specific state $\ket{n}$ and the other in a different (orthogonal) specific state $\ket{n'}$ with $n'\neq n$. For classical or non-identical particles there are two such states, for indistinguishable identical bosons there is just one. 
(The latter statement applies to indistinguishable identical fermions as well, but we first focus on bosons here.)
One consequence is that non-interacting classical particles are uncorrelated in thermal equilibrium, whereas indistinguishable non-interacting bosons are correlated: they ``bunch,'' with a larger probability to be found in the same state.
But how do {\em non}-interacting particles become correlated?
\subsection{Identical vs indistinguishable}
Consider a single boson and suppose $N$ different (orthogonal) quantum states are accessible to it.  
That is, we assume that, given certain constraints (for example, confinement to a finite volume and fixed values for energy and angular momentum) there is an $N$-dimensional Hilbert subspace that contains all states satisfying those constraints.

Now consider two identical bosons, i.e., bosons whose intrinsic properties (like rest mass, charge, and spin) are the same.
For two non-interacting identical bosons the two sets of accessible single-particle states may be the same (when the bosons satisfy the same constraints) or disjoint (when they satisfy mutually exclusive constraints, e.g., when they are located in different parts of space).
In the former case we call the bosons indistinguishable, in the latter case we call them distinguishable \footnote{Partial distinguishability may be defined in terms of partially overlapping sets of accessible states, or in terms of sets of partially overlapping accessible states.}. And so ``identical'' and ``indistinguishable'' are different concepts (see, e.g., Refs.~\cite{de1975,dieks1,dieks2}).

The importance of identical particles having access to the same set of states for quantum statistics to be relevant, and the importance, therefore, of identical particles actually having the possibility to swap locations was pointed out explicitly in
\cite{leggett2006}. Here and in the companion paper \footnote{Companion paper: S.J. van Enk, {\em Exchanging two identical particles}, {\tt arXiv:1810.05208}} we add some more reasons to emphasize these points.

\subsection{Unitary evolution}\label{UE}
Suppose for a single boson the time evolution over some fixed time interval  is unitary and determined by a unitary matrix $U$ with matrix elements $U_{nm}$ such that
\be
\ket{n}\mapsto \sum_m U_{nm}\ket{m},
\ee
where we have chosen an orthonormal basis $\{\ket{n}\}$ for $n=1\ldots N$ for the accessible subspace. The sum ranges over $m=1\ldots N$. 

If we start two identical non-interacting bosons in the same state $\ket{n}$, then their joint state evolves as
\bea\label{Unm}
\ket{n}_1\ket{n}_2\mapsto 
\sum_{m} U_{nm}\ket{m}_1\otimes
\sum_{m'}U_{nm'}\ket{m'}_2.
\eea
We labeled here the two single-particle Hilbert spaces by `1' and `2'. These are mathematical labels for two Hilbert spaces. (They are not physical labels that could be used to distinguish the two particles \cite{dieks1,dieks2}.)

We may rewrite the right-hand side of (\ref{Unm}) as
\bea\label{Unm2}
\ket{n}_1\ket{n}_2&\mapsto& \sum_m U^2_{nm}\ket{m}_1\ket{m}_2
+\nonumber\\
&&\sum_{m<m'}
\sqrt{2}U_{nm}U_{nm'}
\frac{\ket{m}_1\otimes\ket{m'}_2+\ket{m'}_1\otimes\ket{m}_2}{\sqrt{2}},\nonumber\\
\eea
which 
shows the state is automatically properly symmetrized.
The first line shows that the
amplitude for both bosons to end up in the same state $\ket{m}$ is
\be\label{uncor}
T_{nn\mapsto mm}=(U_{nm})^2.
\ee
The probability to end up in the same state $\ket{m}$ is, therefore, simply the product of the probabilities for each individual boson to end up in that state. That is, there is {\em no} correlation here and this is  not going to give us an explanation  for how non-interacting bosons become correlated. 

On the other hand, if the bosons start out in different states, one in $\ket{n}$, the other in $\ket{n'}$ with $n'\neq n$, the transition amplitude for ending up in the same state $\ket{m}$ is
\be\label{bunch}
T_{nn'\mapsto mm}=\sqrt{2} U_{nm}U_{n'm}.
\ee
This is so because the initial state must be written in the symmetrized form
\be\label{b}
\ket{\Psi(0)}=\frac{\ket{n}_1\otimes\ket{n'}_2+\ket{n'}_1\otimes\ket{n}_2}{\sqrt{2}}
\ee
rather than simply as $\ket{n}_1\otimes\ket{n'}_2$, which would be appropriate for distinct particles.
Each term in the numerator of (\ref{b}) evolves into the {\em same} state $U_{nm}U_{n'm}\ket{m}_1\ket{m}_2$, thus yielding the result (\ref{bunch}).
We see that the amplitude for the bosons to end up in the same state gets an extra factor of $\sqrt{2}$ as compared to the result for two classical (or non-identical) particles.
This gives us the bunching effect we expect for indistinguishable identical bosons: {\em the probability to end up in the same state is twice what it would be for two classical particles.} 

All this is well-known, of course.
The more general statement that the probability for $M$ indistinguishable bosons to end up in the same state
increases by a factor of $M!$ as compared to the classical case can be found as Eq.~(4.21) in Volume III of Feynman's lectures on physics.
It is important to note, however, that this statement applies only to bosons starting in $M$ different (orthogonal) states.

(For fermions we could phrase the conclusion in similar terms: {\em the probability to end up in the same state is zero times what it would be for classical particles.} This, of course, is just the Pauli exclusion principle written in a silly form. In this case the proviso that the fermions start in different states isn't needed.)
\subsection{Hong-Ou-Mandel effect}
\label{HOM}
To illustrate the results of the preceding subsection,
consider a simple example of two noninteracting bosons, namely, two photons impinging on a 50/50 beamsplitter (a device with 2 input ports and 2 output ports). The results quoted below are derived using the second-quantization formalism (which does not make use of the unphysical labels `1' and `2') in the Appendix.

It turns out that if two photons enter the same input port, 
the output contains with 50\% probability one photon in each output port, and with 50\% probability the photons exit in the same output port. That is, the photons behave independently and each is randomly exiting one of the two output ports, just as if they were classical particles (in agreement with Eq.~(\ref{uncor})).

If, on the other hand, each photon enters a different input port, then we get the celebrated Hong-Ou-Mandel effect \cite{hong1987}: both photons will emerge together in one output port, and which port that is, is random.
Now note that the probability to bunch, 100\%, is indeed simply twice the probability of 50\% to exit together if the photons were classical independent particles (in agreement with Eq.~(\ref{bunch})).
So, the Hong-Ou-Mandel effect provides one  example of the simple general bunching rule for pairs of bosons. 
\subsection{Towards thermal equilibrium}
Just having photons interact with ideal 50/50 beamsplitters won't drive them to thermal equilibrium. Just having two bosons undergo unitary evolution won't do that either. What is needed is an irreversible interaction with the environment. For example, an interaction that tends to make the density matrix of the two bosons diagonal in one particular basis that is ``complementary'' to the energy eigenbasis---i.e., such that each basis state has some overlap with every accessible energy eigenstate and vice versa--- will do. A spatially localized basis tends to have that property.  The combination of two competing evolutions, one diagonal in the energy basis and the other diagonal in a complementary basis, has just one possible steady state within the space of fixed energy that is diagonal in both: the micro-canonical ensemble.

Note that we did indeed sneak in an irreversible interaction in the earlier discussion in subsection \ref{UE}: the ``probability for two bosons to end up in the same state'' really refers to the probability of obtaining such a result in a joint measurement. It is the measurement that is irreversible. It is also the mechanism by which the two bosons interact with each other {\em indirectly}, namely by both of them interacting with the same (measurement) device. In addition, note that the statements made in subsection \ref{UE} about particles ending up in the same state, are actually basis-dependent (the statements are about the basis $\{\ket{n}\}$).

It may be useful to see explicitly how combining 
evolution from two types of initial states, one developing no correlations at all, the other multiplying the probability of bunching by a factor of 2,  leads to just the right statistics in equilibrium. Let us consider a very simple case: pairs of photons impinging on a 50/50 beam splitter and then undergoing a measurement
that checks merely whether the two photons are in the same output port (we do not care which one) or in different output ports. We keep track of only these two possibilities here.  We can then summarize the evolution due to one beamsplitter/measurement interaction in terms of a $2\times2$ matrix acting on a vector containing the two corresponding probabilities as
\bea\label{map}
\left(
\begin{array}{c}
P_{{\rm same}} \\
P_{{\rm diff}}     
\end{array}
\right)
\longmapsto
\left(
\begin{array}{cc}
1/2 & 1     \\
1/2 &       0
\end{array}
\right)
\left(
\begin{array}{c}
P_{{\rm same}} \\
P_{{\rm diff}}     
\end{array}
\right).
\eea
As is easily verified, this map has the correct (Bose-Einstein) steady-state solution
\be\label{ss}
\left(
\begin{array}{c}
P_{{\rm same}} \\
P_{{\rm diff}}     
\end{array}
\right)_{{\rm steady state}}=\left(
\begin{array}{c}
2/3 \\
1/3   
\end{array}
\right).
\ee
How the correct steady state emerges can be seen  by iterating the map (\ref{map}): for example, just 10 beamsplitter/measurement interactions produce the evolution matrix
\bea
\left(
\begin{array}{cc}
1/2 & 1     \\
1/2 &       0
\end{array}
\right)^{10}\approx
\left(
\begin{array}{cc}
0.667 & 0.666     \\
0.333 &       0.334
\end{array}
\right),
\eea
thus driving any initial probability distribution very close to the thermal distribution (\ref{ss}) in just ten steps.

\subsection{Conclusions}

Correlations between identical particles arise only from initial states in which they occupy {\em different} states. Moreover,  non-interacting particles (i.e., particles that don't interact with each other) 
do need to interact with the same environment in order to thermalize. 
Two different types of evolution are needed for thermalization, one diagonal in the energy basis (Hamiltonian evolution, which is always present), the other irreversible and diagonal in a complementary basis (for example, a joint spatially localized measurement interaction).

We considered just two identical particles here. The above conclusions about thermalization apply to more than two particles as well.  What does change notably when going to more than two particles is that the patterns of destructive interference between multiple paths leading to the same final state \cite{tichy2012} become much more complex. The complexity of such interference effects for indistinguishable bosons underlies the difficulty of simulating {\em boson sampling} \cite{aaronson2011,harrow2017} on a classical computer.
\subsection*{Appendix}
The standard way to use the second quantization formalism [which does not make use of the (mathematical, not physical) labels for single-particle Hilbert spaces that were used in the main text]
for deriving the Hong-Ou-Mandel effect \cite{hong1987} starts by introducing two input modes $a$ and $b$ (one for each input port of the beamsplitter) and two output modes $a'$ and $b'$, which are described by creation and annihilation operators that satisfy the appropriate bosonic commutation relations.
Denoting the two input creation operators by
$\had$ and $\hbd$ and the output creation operators by $\hadp$ and $\hbdp$ we have the following general relations between input and output states.

An input state with $N$ photons in input mode $a$ and $M$ photons in input mode $b$ is given by
\be
\ket{\psi}_{{\rm in}}=\frac{(\had)^N}{\sqrt{N!}}\frac{(\hbd)^M}{\sqrt{M!}}\vac,
\ee
which is transformed by a beamsplitter with (real) transmission and reflection amplitudes $t$ and $r$
to an output state
\be
\ket{\psi}_{{\rm out}}=\frac{(\hadp)^N}{\sqrt{N!}}\frac{(\hbdp)^M}{\sqrt{M!}}\vac,
\ee
where
\bea
\hadp&=&t\had+ir\hbd,\nonumber\\
\hbdp&=&t\hbd+ir\had.
\eea
For $N=M=1$ this yields the output state
\bea
\ket{\psi}_{{\rm out}}&=&\hadp\hbdp\vac
\nonumber\\
&=&itr ((\had)^2+(\hbd)^2)\vac
+(t^2-r^2) \had\hbd\vac  \nonumber\\
\eea
For a 50/50 beamsplitter we can set $t=r=1/\sqrt{2}$, and thus obtain the output state
\bea
\ket{\psi}_{{\rm out}}=
i/2 [(\had)^2+(\hbd)^2]\vac,
\eea
which displays the bunching effect discussed in Section \ref{HOM}.
This bunching effect really arises from destructive interference between the two paths  leading to the alternative final state $ \had\hbd\vac$. Note this destructive interference phenomenon generalizes to larger numbers of photons: if $K$ photons enter each input port of the beamsplitter
(so that $N=M=K$ in the above equations), the output is 
\bea
\ket{\psi}_{{\rm out}}=
\frac{\left(i/2[(\had)^2+(\hbd)^2]\right)^K}{K!}\vac,
\eea
in which all combinations containing odd numbers of photons in any mode have canceled out.

\bibliography{bosons2}

\begin{thebibliography}{10}%
\makeatletter
\providecommand \@ifxundefined [1]{%
 \@ifx{#1\undefined}
}%
\providecommand \@ifnum [1]{%
 \ifnum #1\expandafter \@firstoftwo
 \else \expandafter \@secondoftwo
 \fi
}%
\providecommand \@ifx [1]{%
 \ifx #1\expandafter \@firstoftwo
 \else \expandafter \@secondoftwo
 \fi
}%
\providecommand \natexlab [1]{#1}%
\providecommand \enquote  [1]{``#1''}%
\providecommand \bibnamefont  [1]{#1}%
\providecommand \bibfnamefont [1]{#1}%
\providecommand \citenamefont [1]{#1}%
\providecommand \href@noop [0]{\@secondoftwo}%
\providecommand \href [0]{\begingroup \@sanitize@url \@href}%
\providecommand \@href[1]{\@@startlink{#1}\@@href}%
\providecommand \@@href[1]{\endgroup#1\@@endlink}%
\providecommand \@sanitize@url [0]{\catcode `\\12\catcode `\$12\catcode
  `\&12\catcode `\#12\catcode `\^12\catcode `\_12\catcode `\%12\relax}%
\providecommand \@@startlink[1]{}%
\providecommand \@@endlink[0]{}%
\providecommand \url  [0]{\begingroup\@sanitize@url \@url }%
\providecommand \@url [1]{\endgroup\@href {#1}{\urlprefix }}%
\providecommand \urlprefix  [0]{URL }%
\providecommand \Eprint [0]{\href }%
\providecommand \doibase [0]{http://dx.doi.org/}%
\providecommand \selectlanguage [0]{\@gobble}%
\providecommand \bibinfo  [0]{\@secondoftwo}%
\providecommand \bibfield  [0]{\@secondoftwo}%
\providecommand \translation [1]{[#1]}%
\providecommand \BibitemOpen [0]{}%
\providecommand \bibitemStop [0]{}%
\providecommand \bibitemNoStop [0]{.\EOS\space}%
\providecommand \EOS [0]{\spacefactor3000\relax}%
\providecommand \BibitemShut  [1]{\csname bibitem#1\endcsname}%
\let\auto@bib@innerbib\@empty
\bibitem [{Note1()}]{Note1}%
  \BibitemOpen
  \bibinfo {note} {Partial distinguishability may be defined in terms of
  partially overlapping sets of accessible states, or in terms of sets of
  partially overlapping accessible states.}\BibitemShut {Stop}%
\bibitem [{\citenamefont {De~Muynck}(1975)}]{de1975}%
  \BibitemOpen
  \bibfield  {author} {\bibinfo {author} {\bibfnamefont {W.}~\bibnamefont
  {De~Muynck}},\ }\href@noop {} {\bibfield  {journal} {\bibinfo  {journal}
  {Int. J. Theor. Phys.}\ }\textbf {\bibinfo {volume} {14}},\ \bibinfo {pages}
  {327} (\bibinfo {year} {1975})}\BibitemShut {NoStop}%
\bibitem [{\citenamefont {Dieks}(1990)}]{dieks1}%
  \BibitemOpen
  \bibfield  {author} {\bibinfo {author} {\bibfnamefont {D.}~\bibnamefont
  {Dieks}},\ }\href@noop {} {\bibfield  {journal} {\bibinfo  {journal}
  {Synthese}\ }\textbf {\bibinfo {volume} {82}},\ \bibinfo {pages} {127}
  (\bibinfo {year} {1990})}\BibitemShut {NoStop}%
\bibitem [{\citenamefont {Dieks}\ and\ \citenamefont
  {Lubberdink}(2011)}]{dieks2}%
  \BibitemOpen
  \bibfield  {author} {\bibinfo {author} {\bibfnamefont {D.}~\bibnamefont
  {Dieks}}\ and\ \bibinfo {author} {\bibfnamefont {A.}~\bibnamefont
  {Lubberdink}},\ }\href@noop {} {\bibfield  {journal} {\bibinfo  {journal}
  {Found. Phys.}\ }\textbf {\bibinfo {volume} {41}},\ \bibinfo {pages} {1051}
  (\bibinfo {year} {2011})}\BibitemShut {NoStop}%
\bibitem [{\citenamefont {Leggett}(2006)}]{leggett2006}%
  \BibitemOpen
  \bibfield  {author} {\bibinfo {author} {\bibfnamefont {A.~J.}\ \bibnamefont
  {Leggett}},\ }\href@noop {} {\emph {\bibinfo {title} {Quantum liquids: Bose
  condensation and Cooper pairing in condensed-matter systems}}}\ (\bibinfo
  {publisher} {Oxford university press},\ \bibinfo {year} {2006})\BibitemShut
  {NoStop}%
\bibitem [{Note2()}]{Note2}%
  \BibitemOpen
  \bibinfo {note} {Companion paper: S.J. van Enk, {\protect \em Exchanging two
  identical particles}, {\protect \tt arXiv:1810.05208}}\BibitemShut {NoStop}%
\bibitem [{\citenamefont {Hong}\ \emph {et~al.}(1987)\citenamefont {Hong},
  \citenamefont {Ou},\ and\ \citenamefont {Mandel}}]{hong1987}%
  \BibitemOpen
  \bibfield  {author} {\bibinfo {author} {\bibfnamefont {C.-K.}\ \bibnamefont
  {Hong}}, \bibinfo {author} {\bibfnamefont {Z.-Y.}\ \bibnamefont {Ou}}, \ and\
  \bibinfo {author} {\bibfnamefont {L.}~\bibnamefont {Mandel}},\ }\href@noop {}
  {\bibfield  {journal} {\bibinfo  {journal} {Phys. Rev. Lett.}\ }\textbf
  {\bibinfo {volume} {59}},\ \bibinfo {pages} {2044} (\bibinfo {year}
  {1987})}\BibitemShut {NoStop}%
\bibitem [{\citenamefont {Tichy}\ \emph {et~al.}(2012)\citenamefont {Tichy},
  \citenamefont {Tiersch}, \citenamefont {Mintert},\ and\ \citenamefont
  {Buchleitner}}]{tichy2012}%
  \BibitemOpen
  \bibfield  {author} {\bibinfo {author} {\bibfnamefont {M.~C.}\ \bibnamefont
  {Tichy}}, \bibinfo {author} {\bibfnamefont {M.}~\bibnamefont {Tiersch}},
  \bibinfo {author} {\bibfnamefont {F.}~\bibnamefont {Mintert}}, \ and\
  \bibinfo {author} {\bibfnamefont {A.}~\bibnamefont {Buchleitner}},\
  }\href@noop {} {\bibfield  {journal} {\bibinfo  {journal} {New J. Phys.}\
  }\textbf {\bibinfo {volume} {14}},\ \bibinfo {pages} {093015} (\bibinfo
  {year} {2012})}\BibitemShut {NoStop}%
\bibitem [{\citenamefont {Aaronson}\ and\ \citenamefont
  {Arkhipov}(2011)}]{aaronson2011}%
  \BibitemOpen
  \bibfield  {author} {\bibinfo {author} {\bibfnamefont {S.}~\bibnamefont
  {Aaronson}}\ and\ \bibinfo {author} {\bibfnamefont {A.}~\bibnamefont
  {Arkhipov}},\ }in\ \href@noop {} {\emph {\bibinfo {booktitle} {Proceedings of
  the forty-third annual ACM symposium on Theory of computing}}}\ (\bibinfo
  {organization} {ACM},\ \bibinfo {year} {2011})\ pp.\ \bibinfo {pages}
  {333--342}\BibitemShut {NoStop}%
\bibitem [{\citenamefont {Harrow}\ and\ \citenamefont
  {Montanaro}(2017)}]{harrow2017}%
  \BibitemOpen
  \bibfield  {author} {\bibinfo {author} {\bibfnamefont {A.~W.}\ \bibnamefont
  {Harrow}}\ and\ \bibinfo {author} {\bibfnamefont {A.}~\bibnamefont
  {Montanaro}},\ }\href@noop {} {\bibfield  {journal} {\bibinfo  {journal}
  {Nature}\ }\textbf {\bibinfo {volume} {549}},\ \bibinfo {pages} {203}
  (\bibinfo {year} {2017})}\BibitemShut {NoStop}%
\end{thebibliography}%

\end{document}